\documentclass[prx,twocolumn,floatfix,superscriptaddress]{revtex4-1}
\usepackage{inputenc}
\usepackage{amsmath}
\usepackage{graphicx}
\usepackage[usenames]{color} 
\usepackage{amssymb}
\usepackage[toc,page,titletoc]{appendix}
\usepackage{amsmath}
\usepackage{amsfonts}
\usepackage{amsthm}
\usepackage{enumitem}
\usepackage{qcircuit}
\usepackage{subfigure}
\usepackage{soul}
\usepackage[english]{babel}

\usepackage{lineno}
\usepackage{hyperref}
\hypersetup{
    colorlinks = true,
    urlcolor = {black},
    citecolor = {green},
    linkcolor = {red}
}
\usepackage{letltxmacro}
\LetLtxMacro{\ORIGselectlanguage}{\selectlanguage}
\makeatletter
\DeclareRobustCommand{\selectlanguage}[1]{%
  \@ifundefined{alias@\string#1}
    {\ORIGselectlanguage{#1}}
    {\begingroup\edef\x{\endgroup
       \noexpand\ORIGselectlanguage{\@nameuse{alias@#1}}}\x}%
}
\newcommand{\definelanguagealias}[2]{%
  \@namedef{alias@#1}{#2}%
}
\makeatother

\definelanguagealias{en}{english}

\newcommand{\beq}{\begin{equation}}
\newcommand{\eeq}{\end{equation}}
\def\la{{\langle}}
\def\ra{{\rangle}}

\newcommand{\ket}[1]{|#1\ra}
\newcommand{\bra}[1]{\la #1|}
\newcommand{\braket}[2]{\la #1|#2 \ra}
\newcommand{\ie}{\textit{i.e.,} }

\def\la{{\langle}}
\def\ra{{\rangle}}

\begin{document}

\setlength\parskip{0.5mm}

\title{Measuring the tangle of three-qubit states}
\author{Adri\'an P\'erez-Salinas}
\affiliation{Departament de F\'isica Qu\`antica i Astrof\'isica and Institut de Ci\`encies del Cosmos (ICCUB), Universitat de Barcelona, Mart\'i i Franqu\`es 1, 08028 Barcelona, Spain.}
\affiliation{Barcelona Supercomputing Center, Barcelona, Spain.}
\author{Diego Garc\'ia-Mart\'in}
\affiliation{Departament de F\'isica Qu\`antica i Astrof\'isica and Institut de Ci\`encies del Cosmos (ICCUB), Universitat de Barcelona, Mart\'i i Franqu\`es 1, 08028 Barcelona, Spain.}
\affiliation{Barcelona Supercomputing Center, Barcelona, Spain.}
\affiliation{Instituto de F\'{i}sica Teórica, UAM-CSIC, Madrid, Spain.}
\author{Carlos Bravo-Prieto}
\affiliation{Departament de F\'isica Qu\`antica i Astrof\'isica and Institut de Ci\`encies del Cosmos (ICCUB), Universitat de Barcelona, Mart\'i i Franqu\`es 1, 08028 Barcelona, Spain.}
\affiliation{Barcelona Supercomputing Center, Barcelona, Spain.}
\author{Jos\'e I. Latorre}
\affiliation{Departament de F\'isica Qu\`antica i Astrof\'isica and Institut de Ci\`encies del Cosmos (ICCUB), Universitat de Barcelona, Mart\'i i Franqu\`es 1, 08028 Barcelona, Spain.}
\affiliation{Center for Quantum Technologies, National University of Singapore, Singapore.}
\affiliation{Technology Innovation Institute, Abu Dhabi, UAE.}

\begin{abstract}
We present a quantum circuit that transforms an unknown three-qubit state into its canonical form, up to relative phases, given many copies of the original state. The circuit is made of three single-qubit parametrized quantum gates, and the optimal values for the parameters are learned in a variational fashion. Once this transformation is achieved, direct measurement of outcome probabilities in the computational basis provides an estimate of the tangle, which quantifies genuine tripartite entanglement. We perform simulations on a set of random states under different noise conditions to asses the validity of the method.\vspace{1.0em}


\end{abstract}
\maketitle
\section{Introduction}
The description of entanglement in a three-qubit system uncovers the subtle and vast problem of classifying and quantifying multipartite entanglement in a reliable way. Although the concept of entanglement is of central importance in the fields of Quantum Information and Computation \cite{qc-nielsen2011}, or in Condensed Matter Physics \cite{condensedmatter-laflorencie2016}, there is no known general theory of entanglement yet. As the number of qubits increases, an exponentially large number of entanglement invariants under local unitaries can be constructed, and different entanglement classes can be distinguished \cite{tangle-dur2000, tangle-verstraete2002, tangle-datta2018}. Furthermore, the possibility of measuring these entanglement quantifiers on actual states seems out of reach for more than a few qubits \cite{AndreasWinter2003}. 

The mainstream approach to deal with multipartite entanglement consists of considering different bipartitions of the system of $n$ qubits and analyze the entanglement that characterizes them. The mathematical tool usually employed is the Singular Value Decomposition, which describes a pure state as a linear combination of product states from the two partitions of the complete system \cite{schmidt-ekert1995}. In turn, the eigenvalues of this decomposition can be used to compute entanglement entropies \cite{qm-vonneumann2018, renyi-entropy}, which are employed to quantify entanglement. For condensed matter systems, the analysis of subsystems of increasing size displays the phenomenon of scaling of the entanglement entropy, often obeying the so-called area law \cite{JI-riera-Review}.

In contradistinction to bipartite states, there is no simple equivalent to the Singular Value Decomposition for tripartite systems \cite{schmidt-peres1995,schmidt-pati2000}. In that case, a canonical representation allows to set several coefficients of the original state to zero and fix some of its relative phases through local unitaries. In particular, the canonical form of three-qubit states was found by Ac\'in et~al. in Reference~\cite{tangle-acin2000}.

When dealing with pure bipartite states, a variational quantum algorithm \cite{larose2019, qsvd-bravo2019} can be trained on several copies of the original state in order to discover the local unitaries that reveal its Schmidt form. Then, direct measurements in the computational basis provide the eigenvalues of the Singular Value Decomposition, which in turn are used to compute entanglement entropies. Here, we shall explore a similar strategy to obtain the canonical form and measure the tangle of three-qubit states. We propose a quantum circuit made of three local unitaries, each acting on one of the qubits. The action of these unitaries cast the state into its canonical form, up to relative phases, and can be determined in a variational way. Once this transformation is achieved, the frequencies of measurement outputs in the computational basis are used to compute the tangle of the three-qubit system, which quantifies genuine tripartite entanglement.

The standard procedure for measuring the tangle of a given quantum state involves performing quantum tomography \cite{tomography-mohseni2008}. Such method requires knowledge of $4^3$ observables, obtained through $3^3$ different measurement settings. In contrast, the algorithm herein proposed only needs one measurement setting, namely measuring in the computational basis, but several copies of the state are demanded for the optimization. Overall, both methods involve a similar number of copies. However, our proposal also returns the canonical form of the state.

The rest of the paper is organized as follows. The tangle of three-qubit states is briefly reviewed in Section \ref{sec:tangle}. Then, the algorithm for measuring the tangle on a quantum computer is presented in Section \ref{sec:algorithm}.  The results of simulations under different noise conditions are shown in Section \ref{sec:results}. Finally, conclusions are drawn in Section \ref{sec:conclusions}.

\section{Tangle in Three-Qubit States}
\label{sec:tangle}
 
Let us focus now in more detail in tripartite entanglement \cite{karol2018}. Consider a three-qubit system where each qubit constitutes a partition, namely, $A$, $B$ and $C$,
\begin{equation}
\ket\psi_{ABC} = \sum_{i,j,k = 0}^1 t_{ijk}\,\ket{ijk}\,,
\end{equation}
where $\{|ijk\ra\}$ are the computational-basis states, and the complex coefficients in the tensor $t_{ijk}$ obey a normalization relation. 
A genuine entanglement measure of a three-qubit system $\ket\psi_{ABC}$ is the tangle \cite{entanglement-coffman2000}, denoted by $\tau$. It can be obtained from Cayley's hyperdeterminant, which is a generalization of a square-matrix determinant \cite{cayley1894}. To be precise,
\begin{equation} \label{tangle}
\tau = 4 \,|\rm{Hdet}(t_{ijk})|\,.
\end{equation}
In this case, the hyperdeterminant $\rm{Hdet}(t_{ijk})$ is a polynomial of order four in the amplitudes $\{t_{ijk}\}$ \cite{hyperdeterminant-gelfand1994},
\begin{equation} \label{Hyperdet}
\begin{array}{l}
{\rm Hdet}(t_{ijk})= \vspace{0.15cm}\\t_{000}^2t_{111}^2+t_{001}^2t_{110}^2+t_{010}^2t_{101}^2+t_{100}^2t_{011}^2                 \vspace{0.1cm} \\
-2(t_{000}t_{111}t_{011}t_{100}+t_{000}t_{111}t_{101}t_{010}+t_{000}t_{111}t_{110}t_{001}+ \vspace{0.1cm}\\ +t_{011}t_{100}t_{101}t_{010}+t_{011}t_{100}t_{110}t_{001}+t_{101}t_{010}t_{110}t_{001}) \vspace{0.1cm}\\ 
+4(t_{000}t_{110}t_{101}t_{011}+t_{111}t_{001}t_{010}t_{100})\,.
\end{array}
\end{equation}

\begin{figure*}[t!]
\centering
\subfigure[\hspace{1mm} Tangle ]{
\includegraphics[width=0.45\linewidth]{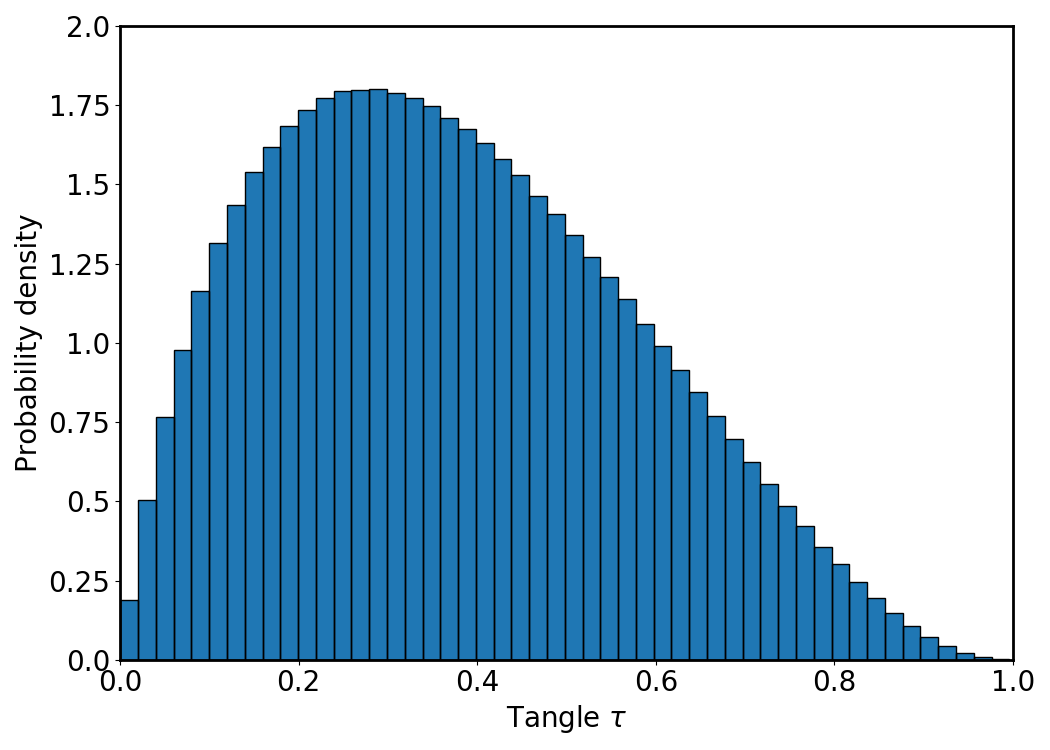}}
\subfigure[\hspace{1mm} Concurrence ]{
\includegraphics[width=0.45\linewidth]{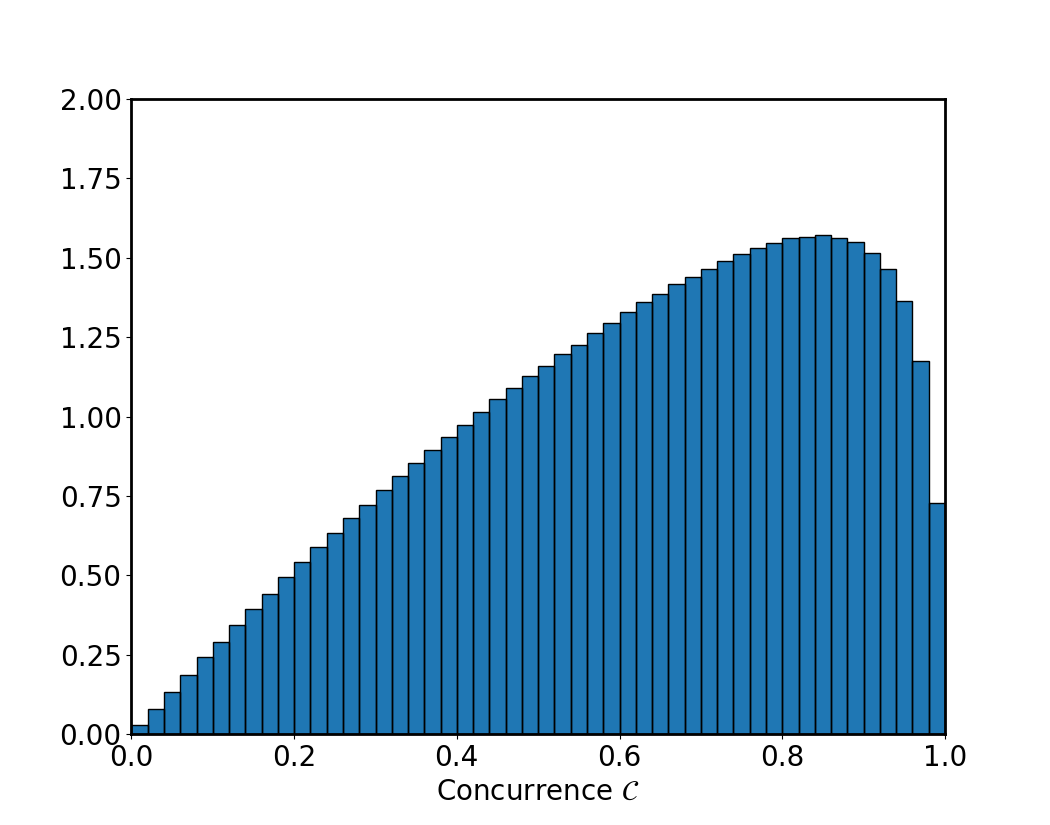}}
\caption{({\bf a}) Probability density of three-qubit random states as a function of the tangle. ({\bf b}) Probability density of two-qubit random states as a function of the concurrence. Three-qubit random states tend to populate values around $\sim 0.3$, while two-qubit random states are mostly distributed at high values.} \label{fig:tangle_distribution}

\end{figure*}

The distribution of the tangle $\tau$ for three-qubit random states is depicted in Figure \ref{fig:tangle_distribution}. We consider random states with $t_{ijk}=a_{ijk} + i\, b_{ijk}$ such that $a_{ijk}$ and $b_{ijk}$ are random real numbers between -0.5 and 0.5, further subject to global normalization. These states tend to populate values of the tangle around $\sim 0.3$. In contrast, the equivalent of the tangle for two-qubit states, namely the concurrence $\mathcal{C} = 2\, |t_{00} t_{11} - t_{01}t_{10}|$, peaks at larger values \cite{entanglement-bengtsson2006}.

In the case of bipartite entanglement, knowledge of the Schmidt coefficients suffices to compute entanglement measures, whereas a full description of a three-qubit state is needed for computing the tangle. However, that being the case, a canonical representation of the three-qubit state may be achieved via local unitaries (LU), allowing for an easier characterization of the entanglement structure. Note that entanglement is not affected by LU \cite{local-kraus2010}. This property of entanglement invariance under local unitary operations is a cornerstone of entanglement theory. 

In this sense, the canonical representation allows to set several amplitudes of the original state to zero and fix some of its relative phases. A canonical form of a tripartite state such that it respects all its entanglement invariants must be constructed with the use of three local unitaries $U_A\otimes U_B \otimes U_C$, each acting on a partition. For a three-qubit state, the complete rationale for this construction goes as follows. The total number of degrees of freedom of a three-qubit state is $ 2 \times 2^3$ real numbers for the coefficients $t_{ijk}$, minus a global phase and norm constraints, which makes a total amount of 14. Now, we remove the freedom carried by the three single-qubit unitaries, which is 3$\times$3. Thus, the number of degrees of freedom is 5. In consequence, there are 5 entanglement invariants under local unitaries \cite{tangle-acin2000}. Note that a similar argument applied to $n$ qubits shows that the number of entanglement invariants grows as $ 2 \times 2^n - 3 n -2$.

It is then always possible to bring a three-qubit state to a canonical form, where three amplitudes are set to zero and only one relative phase remains \cite{tangle-acin2000}. This canonical form reads
\begin{equation}\label{eq:canonical}
\ket{\varphi} = \lambda_0  \ket{000} + \lambda_1 e^{i\phi} \ket{100} + \lambda_2 \ket{101}+ \lambda_3 \ket{110} + \lambda_4 \ket{111} \,,
\end{equation}
where $\{\lambda_i\}$ are real positive values and $\phi$ is a relative phase $0\leq\phi\leq \pi$, attached by convention to $\ket{100}$.
Once the canonical form of the tripartite state is obtained, it is possible to compute the 5 entanglement invariants \cite{tangle-sudbery2001} as
\begin{equation} \label{invariant}
\begin{array}{l}
\frac{1}{2} \leq I_1 \equiv Tr \,\rho_A ^2= 1-2 \mu_0(1-\mu_0-\mu_1)
 \leq 1 \vspace{0.1cm}\\
\frac{1}{2} \leq I_2 \equiv Tr\, \rho_B ^2= 1-2
 \mu_0(1-\mu_0-\mu_1-\mu_2)-2 \Delta
 \leq 1 \vspace{0.1cm}\\
\frac{1}{2} \leq I_3 \equiv Tr\, \rho_C ^2= 1-2
 \mu_0(1-\mu_0-\mu_1-\mu_3)-2 \Delta
 \leq 1 \vspace{0.1cm}\\ 
\frac{1}{4} \leq I_4 \equiv Tr\,( \rho_A \otimes \rho_B \; \rho_{AB})=\\
\phantom{\frac{1}{4} \leq I_4 =\, } 1+\mu_0(\mu_2 \mu_3 -\mu_1 \mu_4 -2\mu_2-3\mu_3-3\mu_4)\vspace{0.1cm}\\
\phantom{\frac{1}{4} \leq I_4 =} -(2-\mu_0)\Delta\leq 1\vspace{0.1cm}\\
0 \leq I_5 \equiv |{\rm Hdet}(t_{ijk})|^2=  \mu_0^2 \mu_4^2 \leq 
 \frac{1}{16}\,, 
\end{array}
\end{equation}
where $\mu_i = \lambda_i^2$ and $\Delta = |\lambda_1\lambda_4 e^{i\phi} - \lambda_2 \lambda_3|^2$. Therefore, from Eq. \eqref{tangle} and Eq. \eqref{invariant} follows that
\begin{equation}
\tau= 4\,\sqrt{I_5} = 4\,\mu_0 \mu_4\,.
\end{equation}
Consequently, given a state in its canonical form, the tangle can be directly computed as the product of the outcome probabilities of the states $\ket{000}$ and $\ket{111}$ in the computational basis, multiplied by four.

\section{Quantum Algorithm for Measuring the Tangle}
\label{sec:algorithm}

\begin{figure}[t]
\[
\Qcircuit @C=0.85em @R=.9em  @!R{
& & & \qw & \gate{U(\vec{\theta}_{A})_{opt}} & \qw & \qw \\
\lstick{\ket \psi_{ABC}} & & &\qw & \gate{U(\vec{\theta}_{B})_{opt}} & \qw & \qw & & & \ket{\tilde{\varphi}} \\
& & & \qw & \gate{U(\vec{\theta}_{C})_{opt}} & \qw & \qw 
\gategroup{1}{2}{3}{2}{.7em}{\{} 
\gategroup{1}{7}{3}{7}{.7em}{\}}
}
\]
\caption{Quantum circuit required for driving an unknown state $\ket\psi_{ABC}$ into its up-to-phases canonical form $\ket{\tilde{\varphi}}$. The optimal parameters $(\vec \theta_A, \vec \theta_B, \vec \theta_
C)_{\rm opt}$ are chosen variationally.}
\label{fig:local_unitary}
\end{figure}
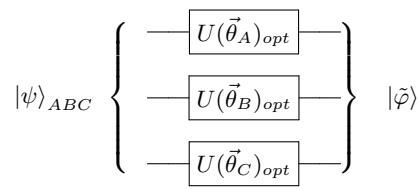

Let us assume that we receive an unknown three-qubit state $\ket\psi_{ABC}$. Our goal is to perform local unitary operations on this state in order to transform it to its canonical form in Equation \eqref{eq:canonical}. Such operations are defined as
\begin{equation}\label{eq:unitary}
\ket{\varphi} = U_A(\vec\theta_A)\otimes U_B(\vec\theta_B)\otimes U_C(\vec\theta_C)\, \ket\psi_{ABC},
\end{equation}
where $\ket\varphi$ is the canonical form of $\ket\psi_{ABC}$ (we drop the subscript $ABC$ in the canonical form for convenience), and each unitary takes the form
\begin{equation}
U(\vec{\theta}) = \begin{pmatrix}
\cos\theta_0 / 2 &  -e^{i\theta_1}\sin\theta_0 / 2 \\ e^{i\theta_2} \sin\theta_0 / 2 & e^{i(\theta_1+\theta_2)} \cos\theta_0/2
\end{pmatrix},
\end{equation}
with
$\vec\theta = (\theta_0, \theta_1, \theta_2)$. It is then necessary to find the values $(\vec \theta_A, \vec \theta_B, \vec \theta_C)_{\rm opt}$ that achieve this transformation. We will follow a hybrid variational strategy and define 
\begin{equation}\label{eq:argmin}
(\vec \theta_A, \vec \theta_B, \vec \theta_C)_{\rm opt} = {\rm argmin}\left( \mathcal{C}(\vec \theta_A, \vec \theta_B, \vec \theta_C)\right)\,,
\end{equation}
where $\mathcal{C}$ is the cost function, defined as
\begin{equation}\label{eq:cost_function}
\mathcal{C}(\vec \theta_A, \vec \theta_B, \vec \theta_C) =  \sum_i \,|\bra{i\,}\, U(\vec \theta_A, \vec \theta_B, \vec \theta_C) \ket{\psi_{ABC}} \,|^2,
\end{equation}
where $i\in\{001,010,011\}$. Notice that the optimal solution, \ie the configuration $(\vec \theta_A, \vec \theta_B, \vec \theta_C)_{\rm opt}$ that renders this cost function equal to zero, transforms $\ket\psi_{ABC}$ into an up-to-phases canonical form $\ket{\tilde{\varphi}}$, given by
\begin{equation}
\begin{split}
\ket{\tilde{\varphi}} = \lambda_0 \ket{000} +  \lambda_1 e^{i\phi_1}\ket{100} +  \lambda_2 e^{i\phi_2}\ket{101} + \\
+ \lambda_3 e^{i\phi_3}\ket{110} + \lambda_4 e^{i\phi_4}\ket{111}\,.
\end{split}
\end{equation}
Such transformation is less restrictive than the canonical transformation in Equation \eqref{eq:unitary}. Therefore, there exist many possible optimal parameters. The quantum circuit implementing this operation is depicted in Figure \ref{fig:local_unitary}.  
Once the optimal parameters are obtained, it is straightforward to measure the tangle $\tau$ in an actual quantum computer. This quantity will be equal to 
\begin{equation}
\tau = 4 \,|\braket{000}{\tilde{\varphi}}\braket{111}{\tilde{\varphi}}|^2 = 4 \,P_{000} P_{111}\,,
\end{equation}
where $P_{ijk}$ is the probability of measuring $\ket{ijk}$. The statistical additive error of $P_{ijk}$ is given by the sampling process of a multinomial distribution, that is, $\sqrt{P_{ijk}(1 - P_{ijk}) / M}$, where $M$ is the number of measurements.

We propose a manner to mitigate random errors occurring when computing the tangle, via post-selection. After the optimization is completed, and a low value of the cost function is obtained, it is licit to assume that $\ket\psi_{ABC}$ has been properly transformed into $\ket{\tilde{\varphi}}$. Thus, if the outcome of a measurement is either $\ket{001}, \ket{010} {\rm \, or \,} \ket{011}$ after the transformation into the up-to-phases canonical form, it is due to an error in the circuit. In this case, this outcome can be discarded.

\section{Simulations}
\label{sec:results}
The algorithm for measuring the tangle can be benchmarked on simulations. We considered a set of 1000 random states, accounting for finite sampling and noise. The state $\ket{\rm GHZ} = \left(\ket{000} + \ket{111}\right) / \sqrt{2}$ has been treated as a particular case, as it is the one that maximizes the tangle and, in addition, it is already in its canonical form. To be precise, we sampled $10^4$ times and introduced random Pauli errors in the quantum circuits in every run, for increasing noise levels. Each measure of the tangle has been repeated ten times, with and without post-selection. We employed the standard Python Library {\tt Scipy} \cite{scipy} for the optimization procedure. In particular, we employed the Powell method as it was found to provide accurate results \cite{powell-powell1964}. The mean number of optimization steps is of the order of a few hundred. 

Not all optimization instances were found to be satisfactory. Some trials did not reach a proper minimum during the first attempt. In order to avoid outliers, only those instances whose cost function was under a certain threshold were accepted. For those that did not match this criterium, the algorithm was rerun. A maximum number of five attempts were allowed.

\subsection{Error Model}\label{sec:errors}
We now present the error model that we have used in the simulations. In this model, single-qubit gates can appear randomly with certain probabilities, to be discussed later. These gates modify the state within the quantum circuit and may appear only after applying the unitary gates from Equation \eqref{eq:unitary}. As the algorithm for measuring the tangle does not require the use of entangling gates, we assume that the qubits have no cross-talk, and thus two-qubit errors are omitted. 

We consider two different types of error. First, random bit-flips, phase-flips and bit-phase-flips are modeled with Pauli-$X, Z, {\rm\, and\, }Y$ gates respectively. All of them may appear sequentially for each one of the qubits. The second kind of error is measurement errors, which are modeled with a Pauli-$X$ gate appearing just before readout. A scheme for the occurrence of these gates is shown in Figure \ref{fig:error_model}.

Every gate has an independent probability of appearing in the circuit,  \ie all error events are uncorrelated. Therefore, the probability of one error $\varepsilon$ occurring is
\begin{equation}
    {\rm Prob}_\varepsilon = p_\varepsilon \prod_{e \neq \varepsilon}(1 - p_e),
\end{equation}
where $p_e$ is the probability that one error occurs, and the product runs over all possible errors. This can be easily extended to calculate the probability of occurrence of a higher number of errors. 

The probabilities of single-qubit and measurement errors are taken as $0.1\, t\%$ and $1\, t\%$ respectively, where $t=\{0, 1, 2, 3, 4, 5\}$ is a tuning parameter. These numbers were selected in agreement with the orders of magnitude present in the experiment in Reference \cite{supremacy-google2019}. Considering $t=5$, there is one error in $\sim 17\%$ of the samples, and there are $\sim 1.5\%$ of events with two errors. Three or more errors are unlikely to happen, with appearance rates under $0.1\%$.

As Pauli-$X, Y, Z$ gates do not commute, choosing to apply first one or another is not equivalent. However, the probability of this kind of events is very low. For instance, the probability of two measurement errors is $p \sim 10^{-4}$, of one measurement and one single-qubit error is $p \sim 10^{-5}$, and for two single-qubit errors is only $p \sim 10^{-6}$. Therefore, we choose one particular ordering. Notice as well that the tangle is not affected by some of these errors, such as Pauli-$Z$ errors alone.

\begin{figure}[t!]
\[
\Qcircuit @C=0.85em @R=.9em  @!R{
 & & & & {\rm Single-qubit} & & & & {\rm Measurement} & & & & & & & \\
 & & \qw & \gate{X} & \gate{Y} & \gate{Z} & \qw & \gate{X} & \meter & \qw \\
\lstick{\ket {\tilde{\varphi}}} & & \qw & \gate{X} & \gate{Y} & \gate{Z} & \qw & \gate{X} & \meter & \qw & & & {\rm Measure}\\
 & & \qw & \gate{X} & \gate{Y} & \gate{Z} & \qw & \gate{X} & \meter & \qw
\gategroup{2}{2}{4}{2}{.7em}{\{}
\gategroup{2}{10}{4}{10}{.7em}{\}}
\gategroup{2}{4}{4}{6}{.7em}{--}
\gategroup{2}{8}{4}{9}{.7em}{--}
}
\]
\caption{Error model for the simulations. Single-qubit and measurements errors can occurr following the scheme of the figure, and may happen with probabilities $0.1 \, t\%$ and $1 \, t\%$, respectively, for $t=\{0, 1, 2, 3, 4, 5\}$. All errors are uncorrelated. This circuit is to be applied after that in Figure \ref{fig:local_unitary}.}
\label{fig:error_model}
\end{figure}
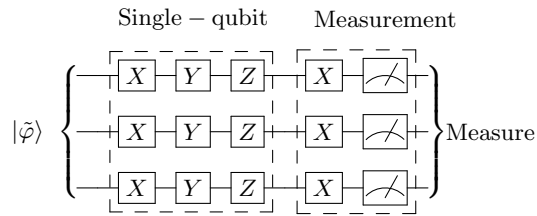
\begin{figure}[t!]
\centering
\includegraphics[width=0.85\linewidth]{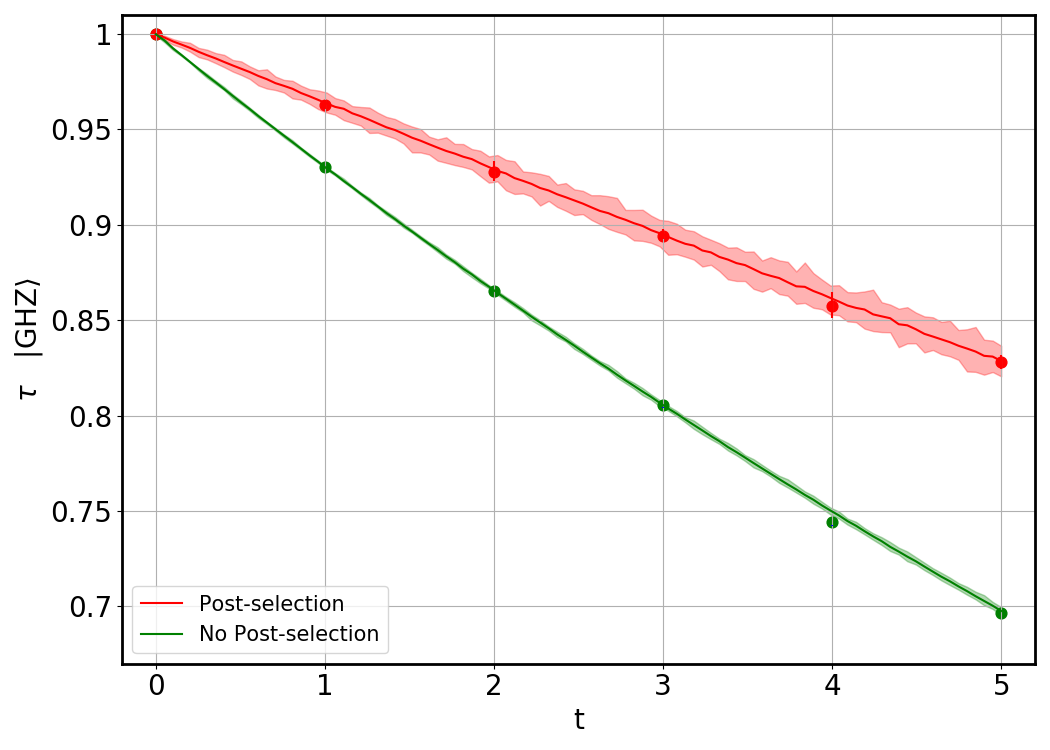}
\caption{Tangle of the $\ket{\rm GHZ}$ state {\sl vs.} parameter $t$ quantifying gate and measurement errors. Solid lines represent averaged results for the tangle obtained without optimization, while the shadowed regions span all results (again without optimization). The dots are the results for the full optimization method applied to the $\ket{\rm GHZ}$ state as if it were an unknown input state. Colors indicate whether post-selection was applied or not. The results indicate that the optimization procedure does not degrade the quality of the estimation of the tangle.}
\label{fig:GHZ}
\end{figure}

\subsection{Results}
\begin{figure*}[t!]
\centering
\subfigure[\hspace{2mm}No error]{
\includegraphics[width=0.85\linewidth]{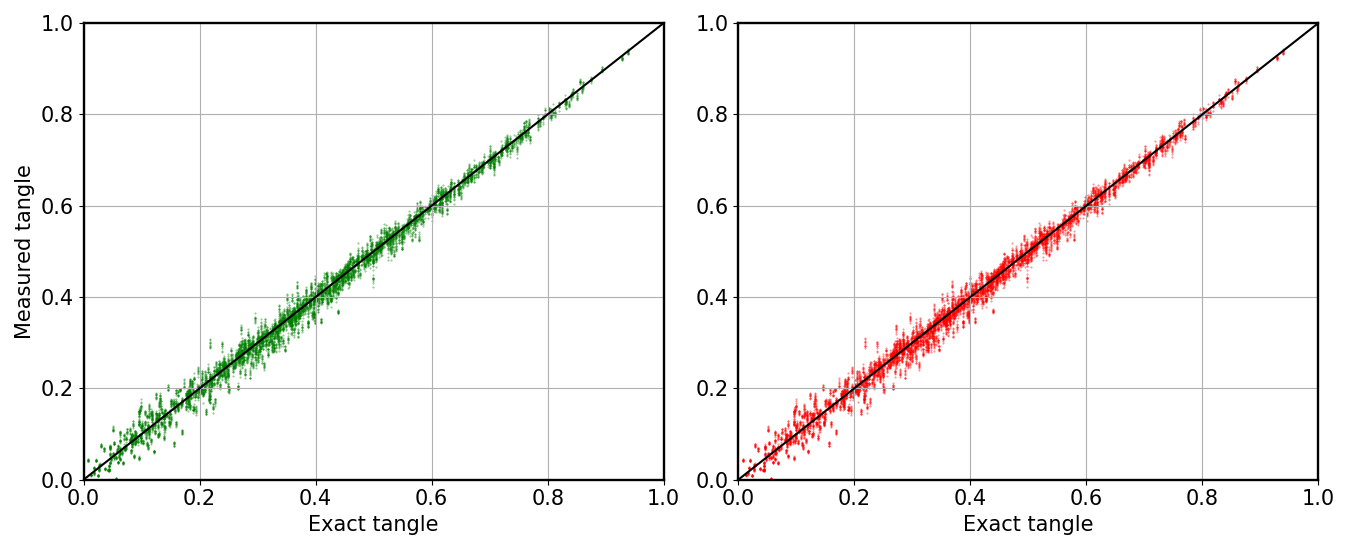}}
\subfigure[\hspace{2mm}Maximum error]{
\includegraphics[width=0.85\linewidth]{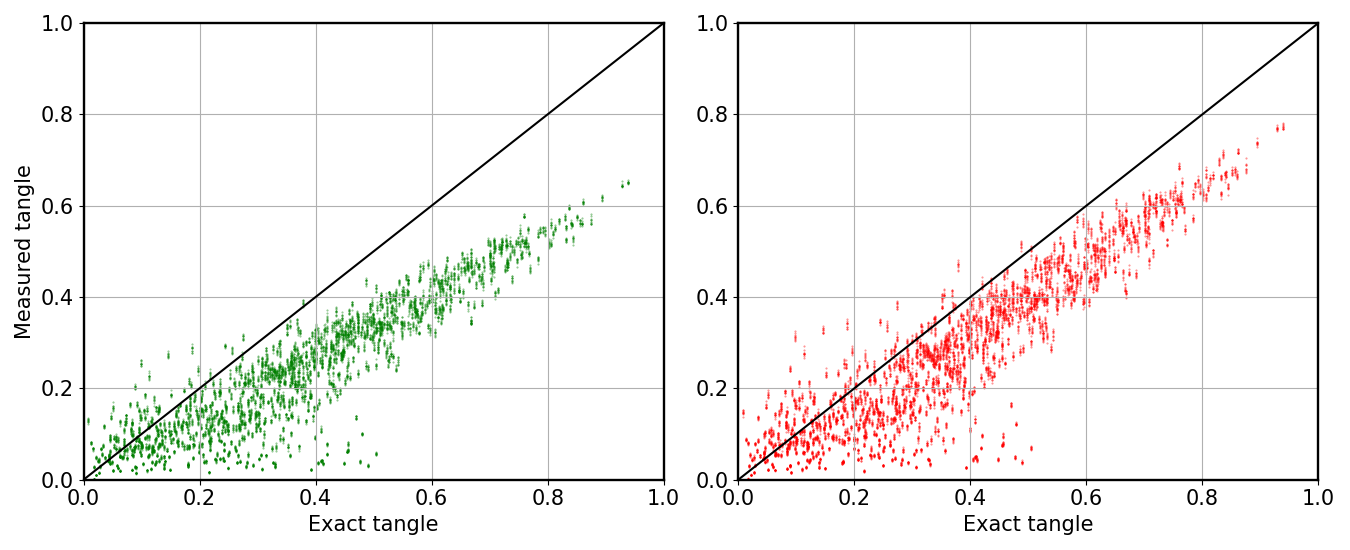}}
\caption{Measured tangle {\sl vs.} exact tangle, for three-qubit random states. Results in green were obtained without applying post-selection, in contrast to those in red. ({\bf a}) Results with no gate errors. ({\bf b}) Results considering the maximum gate error allowed in this paper, t = 5. In all figures, the solid black line represents ideal measurement of the tangle. As the errors decrease, we observe convergence towards the exact tangle.
}
\label{fig:measured_tangles}
\end{figure*}

Two different types of simulations have been carried out. First, we study the $\ket{\rm GHZ}$ state, which maximizes the tangle, $\tau=1$. This state is already in its canonical form. Therefore, there is no need for applying single-qubit gates, and no optimization procedure is needed in this case. The averaged results for the tangle obtained without optimization are represented by solid lines in Figure \ref{fig:GHZ}, while the shadowed regions span all results. On the other hand, the full procedure can be applied to the $\ket{\rm GHZ}$ state as if it were an unknown input state. The results obtained in the latter manner are shown as dots in Figure \ref{fig:GHZ}. This allows us to check that the distortion induced by the optimization procedure does not play a significant role. Both methods were simulated with and without post-selection.

Secondly, we consider 1000 random states and simulate the proposed algorithm for measuring the tangle. The final results are depicted in Figure \ref{fig:measured_tangles}. As it should be expected, results are better for post-selected cases. Those random states whose optimization returned a value of the cost function $\mathcal{C}$ above a certain threshold have been discarded. This process cleans the points far from the solid line in Figure \ref{fig:measured_tangles}. From now, we set said threshold to $ t\, 2\%$, \ie for $t=5$ we allow a minimization error up to $10\%$.

\begin{figure}[t]
\centering
\includegraphics[width=0.85\linewidth]{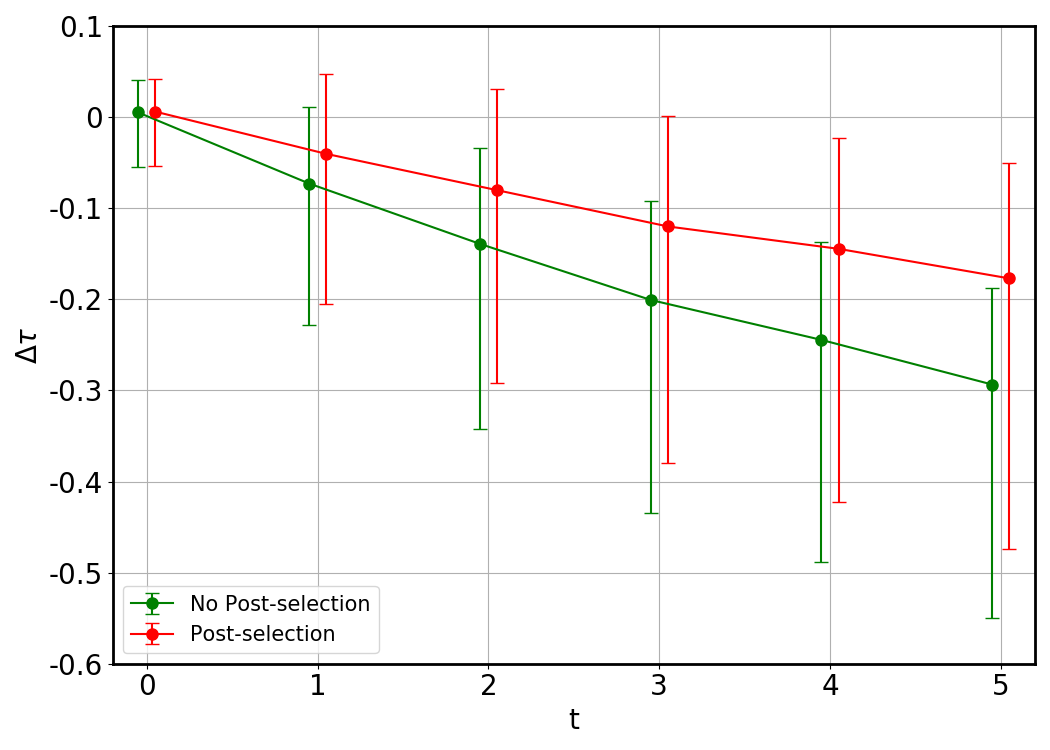}
\caption{Relative error of the tangle of 1000 random states, with a $2\, t\%$ threshold in the cost function value, as a function of the error parameter $t$. Dots correspond to average values, and  error bars span $70\%$ of the measurements. Colors indicate whether post-selection has been applied or not. Note that the algorithm measures the correct tangle in the absence of noise, but tends to underestimate the tangle under its presence.}
\label{fig:relative_errors}
\end{figure}

From the results obtained, it is possible to conclude that circuits with no errors can estimate the tangle properly. In contrast, circuits where errors occur present a tendency to return values of the tangle which are lower than the exact ones. Besides, dispersion increases with the errors.
We present, in Figure \ref{fig:relative_errors}, results of the relative errors in the computation of the tangle, given by 
\begin{equation}
\Delta\tau = \frac{\tau^\prime - \tau}{\tau}\,,
\end{equation}
where $\tau^\prime$ is the estimate obtained through the variational method, and $\tau$ is the exact result. It can be observed that the average behavior of this algorithm in the presence of noise is to underestimate the tangle, as already mentioned. The procedure alone returns an estimate of the tangle $\sim -30\%$ lower than the exact value, while post-processing reduces the error to $\sim -17\%$. 

\section{Conclusions}
\label{sec:conclusions}
We have presented a variational quantum algorithm that casts an unknown three-qubit state into its canonical form, up to relative phases, given many copies of it.  Subsequently, the tangle can be readily measured. The idea behind this procedure is to set three out of eight amplitudes, namely those corresponding to $\ket{001}, \ket{010}$ and $\ket{011}$, to zero. Furthermore, a post-selection scheme allows for a mitigation of the errors.

We have performed simulations on a set of random states to benchmark the proposed algorithm under different noise conditions. We have found that the quantum circuit delivers the correct value of the tangle, with a degradation of the results as the noise levels increase. To be precise, assuming errors comparable to those in state-of-the-art quantum processors, the average relative error is of the order of $\sim -17\%$. It is noteworthy that the tangle is, in most cases, underestimated.

This algorithm does not provide an improvement in the required number of copies of the quantum state, compared to quantum tomography. Nevertheless, the method herein proposed also returns the canonical form of the states and, therefore, might be used as a module in other algorithms. For instance, it can be applied as a pre-processing for a quantum classifier. That is, once the canonical form is cast, the quantum classifier may use this feature to distinguish between different quantum states for a particular task.
\vspace{0.3cm}

{\bf Author contributions:}
Funding acquisition, J.I.L.; Investigation, A.P.-S., D.G.-M., C.B.-P. and J.I.L.; Methodology, A.P.-S.,  D.G.-M., C.B.-P. and J.I.L.; Project administration, J.I.L.; Software, A.P.-S.; Supervision, J.I.L.; Validation, A.P.-S.,  D.G.-M., C.B.-P. and J.I.L.; Writing – original draft, A.P.-S.,  D.G.-M., C.B.-P. and J.I.L.; Writing---review \& editing, A.P.-S.,  D.G.-M., C.B.-P. and J.I.L. All authors have read and approved the final manuscript.
\vspace{0.3cm}

{\bf Funding:}
A.P.-S., D.G.-M., C.B.-P. and J.I.L. are supported by Projects PGC2018-095862-B-C22 and Quantum CAT (001- P-001644). A.P.-S., D.G.-M. and C.B.-P. are also supported by Barcelona Supercomputing Center's project Caixabank Computaci\'on Cu\'antica.
\vspace{0.3cm}

{\bf Conflicts of interest:}
The authors declare no conflict of interest.


\begin{thebibliography}{0}%
\makeatletter
\providecommand \@ifxundefined [1]{%
 \@ifx{#1\undefined}
}%
\providecommand \@ifnum [1]{%
 \ifnum #1\expandafter \@firstoftwo
 \else \expandafter \@secondoftwo
 \fi
}%
\providecommand \@ifx [1]{%
 \ifx #1\expandafter \@firstoftwo
 \else \expandafter \@secondoftwo
 \fi
}%
\providecommand \natexlab [1]{#1}%
\providecommand \enquote  [1]{``#1''}%
\providecommand \bibnamefont  [1]{#1}%
\providecommand \bibfnamefont [1]{#1}%
\providecommand \citenamefont [1]{#1}%
\providecommand \href@noop [0]{\@secondoftwo}%
\providecommand \href [0]{\begingroup \@sanitize@url \@href}%
\providecommand \@href[1]{\@@startlink{#1}\@@href}%
\providecommand \@@href[1]{\endgroup#1\@@endlink}%
\providecommand \@sanitize@url [0]{\catcode `\\12\catcode `\$12\catcode
  `\&12\catcode `\#12\catcode `\^12\catcode `\_12\catcode `\%12\relax}%
\providecommand \@@startlink[1]{}%
\providecommand \@@endlink[0]{}%
\providecommand \url  [0]{\begingroup\@sanitize@url \@url }%
\providecommand \@url [1]{\endgroup\@href {#1}{\urlprefix }}%
\providecommand \urlprefix  [0]{URL }%
\providecommand \Eprint [0]{\href }%
\providecommand \doibase [0]{http://dx.doi.org/}%
\providecommand \selectlanguage [0]{\@gobble}%
\providecommand \bibinfo  [0]{\@secondoftwo}%
\providecommand \bibfield  [0]{\@secondoftwo}%
\providecommand \translation [1]{[#1]}%
\providecommand \BibitemOpen [0]{}%
\providecommand \bibitemStop [0]{}%
\providecommand \bibitemNoStop [0]{.\EOS\space}%
\providecommand \EOS [0]{\spacefactor3000\relax}%
\providecommand \BibitemShut  [1]{\csname bibitem#1\endcsname}%
\let\auto@bib@innerbib\@empty
%</preamble>
\end{thebibliography}%


\vspace{0.3cm}

\begin{thebibliography}{--------}
\providecommand{\natexlab}[1]{#1}

\bibitem[Nielsen and Chuang(2010)]{qc-nielsen2011}
Nielsen, M.A.; Chuang, I.L.
\newblock {\em Quantum Computation and Quantum Information: 10th Anniversary
  Edition}; Cambridge University Press: Cambridge, UK,  2010.
\newblock
  {\href{https://doi.org/10.1017/CBO9780511976667}{\detokenize{[doi]}}}.

\bibitem[Laflorencie(2016)]{condensedmatter-laflorencie2016}
Laflorencie, N.
\newblock Quantum entanglement in condensed matter systems.
\newblock {\em Phys. Rep.} {\bf 2016}, {\em 646},~1–59.
\newblock
  {\href{https://doi.org/10.1016/j.physrep.2016.06.008}{\detokenize{[doi]}}}.

\bibitem[Dür \em{et~al.}(2000)Dür, Vidal, and Cirac]{tangle-dur2000}
Dür, W.; Vidal, G.; Cirac, J.I.
\newblock Three qubits can be entangled in two inequivalent ways.
\newblock {\em Phys. Rev. A} {\bf 2000}, {\em 62}.
\newblock
  {\href{https://doi.org/10.1103/physreva.62.062314}{\detokenize{[doi]}}}.

\bibitem[Verstraete \em{et~al.}(2002)Verstraete, Dehaene, De~Moor, and
  Verschelde]{tangle-verstraete2002}
Verstraete, F.; Dehaene, J.; De~Moor, B.; Verschelde, H.
\newblock Four qubits can be entangled in nine different ways.
\newblock {\em Phys. Rev. A} {\bf 2002}, {\em 65}.
\newblock
  {\href{https://doi.org/10.1103/physreva.65.052112}{\detokenize{[doi]}}}.

\bibitem[Datta \em{et~al.}(2018)Datta, Adhikari, Das, and
  Agrawal]{tangle-datta2018}
Datta, C.; Adhikari, S.; Das, A.; Agrawal, P.
\newblock Distinguishing different classes of entanglement of three-qubit pure
  states.
\newblock {\em Eur. Phys. J. D} {\bf 2018}, {\em 72}.
\newblock
  {\href{https://doi.org/10.1140/epjd/e2018-90199-2}{\detokenize{[doi]}}}.

\bibitem[Leifer \em{et~al.}(2004)Leifer, Linden, and Winter]{AndreasWinter2003}
Leifer, M.S.; Linden, N.; Winter, A.
\newblock Measuring polynomial invariants of multiparty quantum states.
\newblock {\em Phys. Rev. A} {\bf 2004}, {\em 69}.
\newblock
  {\href{https://doi.org/10.1103/physreva.69.052304}{\detokenize{[doi]}}}.

\bibitem[Ekert and Knight(1995)]{schmidt-ekert1995}
Ekert, A.; Knight, P.L.
\newblock Entangled quantum systems and the Schmidt decomposition.
\newblock {\em Am. J. Phys.} {\bf 1995}, {\em 63},~415--423.
\newblock
  {\href{https://doi.org/10.1119/1.17904}{\detokenize{[doi]}}}.

\bibitem[Neumann(2018)]{qm-vonneumann2018}
Neumann, J.
\newblock {\em Mathematical foundations of quantum mechanics}; Princeton University Press: Princeton, NJ, USA, 2018.

\bibitem[R\'enyi(1961)]{renyi-entropy}
R\'enyi, A.
\newblock On Measures of Entropy and Information.
\newblock  Proceedings of the Fourth Berkeley Symposium on Mathematical
  Statistics and Probability, Volume 1: Contributions to the Theory of
  Statistics; University of California Press: Berkeley, CA,USA,  1961; pp.
  547--561.
\newblock {\href{https://projecteuclid.org/euclid.bsmsp/1200512181}{\detokenize{[ref]}}}.

\bibitem[Latorre and Riera(2009)]{JI-riera-Review}
Latorre, J.I.; Riera, A.
\newblock A short review on entanglement in quantum spin systems.
\newblock {\em J. Phys. A. Math. Theor.} {\bf 2009},
  {\em 42},~504002.
\newblock
  {\href{https://doi.org/10.1088/1751-8113/42/50/504002}{\detokenize{[doi]}}}.

\bibitem[Peres(1995)]{schmidt-peres1995}
Peres, A.
\newblock Higher order Schmidt decompositions.
\newblock {\em Phys. Lett. A} {\bf 1995}, {\em 202},~16–17.
\newblock
  {\href{https://doi.org/10.1016/0375-9601(95)00315-t}{\detokenize{[doi]}}}.

\bibitem[Pati(2000)]{schmidt-pati2000}
Pati, A.K.
\newblock Existence of the Schmidt decomposition for tripartite systems.
\newblock {\em Phys. Lett. A} {\bf 2000}, {\em 278},~118--122.
\newblock
  {\href{https://doi.org/10.1016/s0375-9601(00)00767-2}{\detokenize{[doi]}}}.

\bibitem[Ac\'{\i}n \em{et~al.}(2000)Ac\'{\i}n, Andrianov, Costa, Jan\'e,
  Latorre, and Tarrach]{tangle-acin2000}
Ac\'{\i}n, A.; Andrianov, A.; Costa, L.; Jan\'e, E.; Latorre, J.I.; Tarrach, R.
\newblock Generalized Schmidt Decomposition and Classification of
  Three-Quantum-Bit States.
\newblock {\em Phys. Rev. Lett.} {\bf 2000}, {\em 85},~1560--1563.
\newblock
  {\href{https://doi.org/10.1103/PhysRevLett.85.1560}{\detokenize{[doi]}}}.

\bibitem[LaRose \em{et~al.}(2019)LaRose, Tikku, O’Neel-Judy, Cincio, and
  Coles]{larose2019}
LaRose, R.; Tikku, A.; O’Neel-Judy, e.; Cincio, L.; Coles, P.J.
\newblock Variational quantum state diagonalization.
\newblock {\em npj Quantum Inf.} {\bf 2019}, {\em 5}.
\newblock
  \href{https://doi.org/10.1038/s41534-019-0167-6}{\detokenize{[doi]}}.


\bibitem[Bravo-Prieto \em{et~al.}(2019)Bravo-Prieto, Garc\'ia-Mart\'in, and
  Latorre]{qsvd-bravo2019}
Bravo-Prieto, C.; Garc\'ia-Mart\'in, D.; Latorre, J.I.
\newblock Quantum Singular Value Decomposer.
\newblock {\em arXiv} {\bf 2019},arXiv:1905.01353.
\newblock 
  \href{http://arxiv.org/abs/1905.01353}{{\normalfont
  [arXiv]}}.

\bibitem{tomography-mohseni2008}
Mohseni, M., A. T. Rezakhani, and D. A. Lidar. 
\newblock Quantum-Process Tomography: Resource Analysis of Different Strategies. 
\newblock {\em Phys. Rev. A} {\bf 2008}
\newblock   \href{https://doi.org/10.1103/PhysRevA.77.032322}{\detokenize{[doi]}}.

\bibitem[Enr\'iquez \em{et~al.}(2018)Enr\'iquez, Delgado, and
  Zyczkowski]{karol2018}
Enr\'iquez, M.; Delgado, F.; Zyczkowski, K.
\newblock Entanglement of Three-Qubit Random Pure States.
\newblock {\em Entropy} {\bf 2018}, {\em 20},~745.
\newblock
  {\href{https://doi.org/10.3390/e20100745}{\detokenize{[doi]}}}.

\bibitem[Coffman \em{et~al.}(2000)Coffman, Kundu, and
  Wootters]{entanglement-coffman2000}
Coffman, V.; Kundu, J.; Wootters, W.K.
\newblock Distributed entanglement.
\newblock {\em Phys. Rev. A} {\bf 2000}, {\em 61}.
\newblock
  {\href{https://doi.org/10.1103/physreva.61.052306}{\detokenize{[doi]}}}.

\bibitem[Cayley(1894)]{cayley1894}
Cayley, A.
\newblock {\em The collected mathematical papers of {A}rthur {C}ayley}; Vol.~1,
  The Cambridge University Press: Cambridge, UK, 1894.%please confirm.

\bibitem[Gelfand \em{et~al.}(1994)Gelfand, Kapranov, and
  Zelevinsky]{hyperdeterminant-gelfand1994}
Gelfand, I.M.; Kapranov, M.M.; Zelevinsky, A.V.
\newblock {\em Discriminants, Resultants, and Multidimensional Determinants};
  Birkh\"{a}user Boston: Cambridge, MA, USA, 1994. %please confirm.
\newblock
  \href{https://doi.org/10.1007/978-0-8176-4771-1}{\detokenize{[doi]}}.

\bibitem[Bengtsson and Zyczkowski(2006)]{entanglement-bengtsson2006}
Bengtsson, I.; Zyczkowski, K.
\newblock {\em Geometry of Quantum States}; Cambridge University Press: Cambridge, UK,  2006.
\newblock
  {\href{https://doi.org/10.1017/cbo9780511535048}{\detokenize{[doi]}}}.

\bibitem[Kraus(2010)]{local-kraus2010}
Kraus, B.
\newblock Local Unitary Equivalence of Multipartite Pure States.
\newblock {\em Physical Review Letters} {\bf 2010}, {\em 104}.
\newblock
  {\href{https://doi.org/10.1103/physrevlett.104.020504}{\detokenize{[doi]}}}.

\bibitem[Sudbery(2001)]{tangle-sudbery2001}
Sudbery, A.
\newblock On local invariants of pure three-qubit states.
\newblock {\em J. Phys. A Math. Gen.} {\bf 2001}, {\em
  34},~643–652.
\newblock
  {\href{https://doi.org/10.1088/0305-4470/34/3/323}{\detokenize{[doi]}}}.

\bibitem[Virtanen \em{et~al.}(2020)Virtanen et~al.]{scipy}
Virtanen, P.; others %newly added, please confirm.
\newblock SciPy 1.0: Fundamental algorithms for scientific computing in Python.
\newblock {\em Nat. Method.} {\bf 2020}.\newblock
  {\href{http://dx.doi.org/10.1038/s41592-019-0686-2}{\detokenize{[doi]}}}.

\bibitem[Powell(1964)]{powell-powell1964}
Powell, M.J.D.
\newblock An efficient method for finding the minimum of a function of several
  variables without calculating derivatives.
\newblock {\em The Comput. J.} {\bf 1964}, {\em 7},~155--162.
\newblock
  {\href{https://doi.org/10.1093/comjnl/7.2.155}{\detokenize{[doi]}}}.

\bibitem[Arute \em{et~al.}(2019)Arute et~al.]{supremacy-google2019}
Arute, F.; others.
\newblock Quantum supremacy using a programmable superconducting processor.
\newblock {\em Nature} {\bf 2019}, {\em 574},~505--510.
\newblock
  {\href{https://doi.org/10.1038/s41586-019-1666-5}{\detokenize{[doi]}}}.
  
\end{thebibliography}
\end{document}